\documentclass[iop]{emulateapj}
\shorttitle{Globular Clusters in NGC 4449}
\shortauthors{Strader et al.}
\def\etal{{\it et al.}}
\def\kms{\,km~s$^{-1}$}

\def\gsim{\;\rlap{\lower 2.5pt
 \hbox{$\sim$}}\raise 1.5pt\hbox{$>$}\;}
\def\lsim{\;\rlap{\lower 2.5pt
\hbox{$\sim$}}\raise 1.5pt\hbox{$<$}\;}
\usepackage{subfigure}
\journalinfo{To Appear in AJ}

\begin{document}

\title{Old Massive Globular Clusters and the Stellar Halo of the Dwarf Starburst Galaxy NGC 4449}

\author{Jay Strader\altaffilmark{1}, Anil C.~Seth\altaffilmark{1,2}, Nelson Caldwell\altaffilmark{1}}
\email{jstrader@cfa.harvard.edu}

\altaffiltext{1}{Harvard-Smithsonian Center for Astrophysics, Cambridge, MA 02138}
\altaffiltext{2}{University of Utah, Salt Lake City, UT 84112}

\begin{abstract}
We use \emph{Hubble Space Telescope} imaging to show that the nearby dwarf starburst galaxy NGC 4449 has an unusual abundance of luminous red star clusters.  Joint constraints from integrated photometry, low-resolution spectroscopy, dynamical mass-to-light ratios, and resolved color-magnitude diagrams provide evidence that some of these clusters are old globular clusters. Spectroscopic data for two massive clusters suggest intermediate metallicities ([Fe/H] $\sim -1$) and subsolar Mg enhancement ([Mg/Fe] $\sim -0.1$ to $-0.2$). One of these clusters may be the nucleus of a tidally disrupting dwarf galaxy; the other is very massive ($\sim 3\times10^{6} M_{\odot}$). We have also identified a population of remote halo globular clusters. NGC 4449 is consistent with an emerging picture of the ubiquity of stellar halos among dwarf galaxies, and study of its globular clusters may help distinguish between accretion and \emph{in situ} scenarios for such halos.

\end{abstract}


\section{Introduction}

NGC 4449 is one of the nearest starburst galaxies, at a distance of $\sim 3.8$ Mpc (Annibali \etal~2008). Its stellar mass is similar to that of the Large Magellanic Cloud (LMC) but has a much higher star formation rate of $\sim 1 M_{\odot}$ yr$^{-1}$ (McQuinn \etal~2010). Extended complexes of \ion{H}{1} gas and disturbed kinematics have led to the conclusion that NGC 4449 has recently undergone a substantial interaction with another galaxy, although the interaction partner has not been conclusively identified (Hunter \etal~1998).

Annibali \etal~(2008) imaged NGC 4449 with the \emph{Hubble Space Telescope} Advanced Camera for Surveys ($HST$/ACS) and visually identified a number of resolved star clusters. They found a very luminous cluster on the west side of the galaxy and suggested that it could be the nucleus of a disrupting dwarf galaxy. 

A surprising number of other massive clusters are apparent from a quick perusal of the $HST$ images. However, unlike the case for an early-type galaxy, many studies have shown (e.g., Seth \etal~2004; Georgiev \etal~2009) that a large fraction of luminous star clusters in star-forming dwarf galaxies are likely to be young ($< 1$ Gyr) or of intermediate-age (1--5 Gyr).

In this paper we present a photometric and spectroscopic study of star clusters in NGC 4449, focusing on the older clusters, showing that the galaxy has an unusual abundance of old massive star clusters.

\section{Star Clusters from $HST$/ACS Imaging}

\subsection{Photometry}

Annibali \etal~(2008) imaged NGC 4449 in three filters (F435W, F555W, F814W) with two contiguous $HST$/ACS pointings, covering the main body of the galaxy. We term these ``A" and ``B" for the east and west pointings, as shown in Figure 2. We searched these two ACS fields for star clusters. Identifications of clusters of all ages were made by eye on three-color images using an interactive program; the clusters were typically partially resolved into stars. Because of the inhomogeneous background, photometric apertures were also chosen by eye during this process. Table 1 lists our catalog of 106 clusters, including 88 candidates of good quality and 18 less convincing ``possible" clusters. Gelatt \etal~(2001) previously identified candidate clusters within a smaller $HST$/Wide Field Planetary Camera 2 pointing; these IDs are cross-referenced in Table 1 (along with those of Annibali \etal~2011a and Rangelov \etal~2011, as discussed below).  The astrometry in this table has been corrected to align the $HST$ images to the astrometric system of the Sloan Digital Sky Survey (SDSS).

We performed aperture photometry with radii determined by eye for each cluster. The background level was estimated at radii between 2 and 5 times the photometric aperture, using the mean pixel value in 20 different annuli, all with area matched to that of the photometric aperture. The uncertainty in the background is typically the dominant error source, and is estimated from the variance of the values in the different background annuli. Uncertainties in our F555W magnitudes are listed in Table 1, and uncertainties for the other filters are comparable.

Magnitudes are given on the Vega system with the appropriate zeropoints from taken from the ACS website (F435W: 25.793, F555W: 25.744, F814W: 25.536). F555W magnitudes are very close to the standard $V$ band and we use the two interchangeably. A mean aperture correction of 0.05 magnitudes was estimated from bright, isolated clusters and has been applied to all the magnitudes in Table 1. Thus, there may be a small systematic magnitude error for objects with sizes very different from the average; colors should be unaffected. These magnitudes are also corrected for a foreground reddening of $E(B-V) = 0.019$ (Schlegel \etal~1998), as are all magnitudes in this paper. No corrections are made for any internal reddening\footnote{B\"{o}ker \etal~(2001) report 
a value of $E(B-V) = 0.35$ toward the nucleus.}. Due to the wide range of background surface brightnesses, the completeness limit of our cluster search was not uniform. The faintest clusters in our catalog have F555W $\sim 22$; we expect that the catalog is complete for F555W $\la 20$.

In addition to measuring integrated magnitudes, we directly estimated the half-light radius within an aperture twice as large as the photometric aperture. For a subset of our clusters with velocity dispersion estimates from MMT/Hectochelle data (\S 3.2), we made more careful estimates of the structural parameters for use in dynamical mass estimates. These structural parameters were measured by fitting elliptical King profiles, convolved with an empirical point spread function, to the $HST$/ACS F814W data. The resulting half-light radii and concentrations are listed in Table 2. Because of the inhomogeneous background around clusters B13 (see \S5.5) and A14, no satisfactory King model fits could be found, and we use their directly measured half-light radii instead.
 
We note that shortly after the completion of this manuscript, two papers were published that presented photometry of star clusters from the same $HST$/ACS data used in this paper (Annibali \etal~2011a; Rangelov \etal~2011). The goals of these papers have little overlap with the present work (focusing on young clusters and X-ray sources, respectively), so we do not present a detailed comparison of our results with these other papers. The catalogs differ most significantly for the younger and fainter clusters, which is as expected, since the boundaries between clusters and other objects (such as unbound associations or asterisms) are less distinct for younger and less massive clusters.
Our catalog focuses on the older clusters, and all objects in the Annibali \etal~(2011a) catalog brighter than our completeness limit ($V=20$) that we consider to be good candidates for old clusters are also in our catalog. To facilitate catalog comparisons for the interested reader, we have cross-referenced the cluster IDs from Annibali \etal~(2011a) and Rangelov \etal~(2011) in Table 1 for objects in common.
 
\subsection{Cluster Classification}

Figure 1 shows a color-color diagram of all of the star clusters in three luminosity bins. A Padova single stellar population model with Z=0.008 (Marigo \etal~2008) is overplotted for ages of $\leq$1~Gyr (the discussion in Annibali \etal~2008 suggests a current gas phase metallicity of $Z \sim 0.006$) , while a Z=0.004 model is shown for ages $>$1~Gyr. 

The clusters evenly separate into two clumps in this color-color diagram.  A color cut of F555W--F814W $> 0.9$ appears to yield a clean sample of clusters with ages greater than 1 Gyr\footnote{The one exception is B3, which appears to be a very young cluster dominated by red supergiants that falls off of the model grids. Further study of this star cluster could prove fruitful. We exclude it from our analysis of old cluster candidates.}. There are 38 high-quality cluster candidates selected by this color cut within the limited field of the view of the $HST$ observations. Of these 38, 15 have $V < 20$ ($M_V \la -8$), which is a convenient demarcation line for massive clusters, equivalent to $\sim 2 \times10^{5} M_{\odot}$ for $V$-band mass-to-light ratio $M/L_V = 1.5$ (applicable for old globular-like clusters). If indeed many or most of these objects are old globular clusters (GCs), rather than less massive intermediate-age objects, NGC 4449 would possess an unusually rich population of GCs. For example, the Milky Way itself has only $\la 40$ GCs more luminous than $M_V = -8$ but a stellar mass nearly 20 times that of NGC 4449.

In the context of more detailed stellar population analysis, the star cluster luminosity function is discussed in \S 6.1.

\begin{figure}
	\epsscale{1.23}
	\plotone{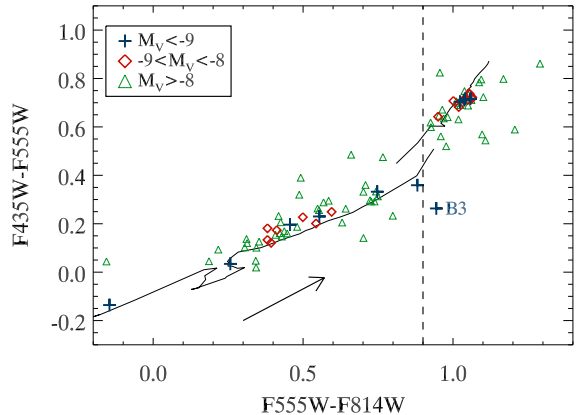}
\figcaption[z_good.eps]{\label{fig:fig_1}
F435W--F555W vs.~F555W--F814W color-color diagram for star clusters in $HST$/ACS imaging of NGC 4449. As indicated in the legend, blue crosses
are the most luminous clusters ($M_V < -9$), red diamonds have $-9 < M_V < -8$, and green triangles the least luminous objects in our sample ($M_V > -8$). The two overplotted solid lines indicate Padova models for ages $\leq$1~Gyr with Z=0.008 (left), and for ages $\geq$1~Gyr with Z=0.004 (right).  The dotted line at F555W--F814W = 0.9 denotes the sample separation between bluer, young clusters and the redder objects with approximate ages $> 1$ Gyr. The unusual cluster B3 is marked. A reddening vector (arrow) with $E(B-V)=0.2$ is also plotted.}
\end{figure}

\section{Spectroscopy}

To study the stellar populations and dynamical masses of clusters in NGC 4449, we obtained low and high-resolution spectroscopy with the MMT. These observations are detailed in the following section.

\subsection{MMT/Hectospec Data}

We targeted clusters for follow-up MMT/Hectospec spectroscopy primarily from the $HST$/ACS catalog discussed above. Hectospec has 300 fibers for both object and sky spectra over a 1 degree field of view (Fabricant \etal~2005). The ACS catalog only covered a small fraction of the Hectospec field of view; in addition, not all ACS sources could be observed because of fiber collisions. As a supplement, we added additional candidate star clusters from the SDSS DR6 (Adelman-McCarthy \etal~2008) imaging of NGC 4449, using the selection criteria $17.0 < V < 20.5$, $(g-r) < 0.8$, $(r-i) < 0.4$, and a projected distance $< 15$\arcmin\ ($\sim 17$ kpc) from the center of the galaxy. These cuts were designed to isolate possible old or intermediate-age star clusters while eliminating the bulk of the foreground contaminants (which, in this magnitude range, are red dwarf stars). The candidates selected from SDSS are discussed further in \S 4.

For the Hectospec observations, we used the 270 l/mm grating, which gave spectral coverage from 3650--9200 \AA \ and a resolution of $\sim$6 \AA. Sky fibers were assigned to random locations over the whole field of view, and we used only those sky spectra that showed no evidence of galaxy light (in emission lines or continuum). The same configuration was observed on two different nights in May 2009 for a combined exposure time of 125 min. The data were reduced as described in Caldwell \etal~(2009).

We derived heliocentric radial velocities through cross-correlation over the wavelength range 3800--6800 \AA\ with a library of templates constructed from star clusters in M31, spanning a range of spectral types. Many of the clusters located within the main body of the galaxy were contaminated by emission; for these objects, we restricted the cross-correlation to the range 3800--4800 \AA\, in which there was less emission. The final radial velocities and errors are listed in Table 3.

Unfortunately, the signal-to-noise ($S/N$) of the Hectospec data was too low for detailed line-index stellar population analysis of most of our targets. Such analysis for two luminous clusters, B13 and B15, is provided in \S 5.1, along with metallicity estimates for a small sample of other old clusters.

In an effort to separate possible intermediate-age clusters from older objects, all of the spectra were visually compared to representative Hectospec data for M31 star clusters, including several reasonably metal-rich clusters ([Fe/H] $> -1$) with ages $\sim 1-2$ Gyr.  Clusters in this age and metallicity range are readily identified through a combination of Balmer and metal lines of moderate strength. No comparable objects were observed among the NGC 4449 spectra. In the cases with sufficient S/N, the clusters identified as young ($< 1$ Gyr) by their optical colors had spectra consistent with this classification, ranging from emission lines (suggesting ages of tens of Myr) to the strong Balmer lines typical of clusters with ages of $\sim 0.2-0.5$ Gyr.

\subsection{MMT/Hectochelle Data}

Our target catalog for MMT/Hectochelle was very similar to that for Hectospec (see previous subsection), with a greater focus on the most luminous candidates. Hectochelle is a high-resolution, single order multi-object spectrograph ($R\sim34000$) with 240 fibers available for assignment, some of which must be used for sky subtraction. All data was taken in the RV31 order, centered at 5215 \AA, which covers the Mg$b$ triplet and is $\sim 150$ \AA\ wide. The spectral data reduction was identical to that described in Caldwell \etal~(2009) and Strader \etal~(2011), producing calibrated one-dimensional spectra for analysis.

We derived heliocentric radial velocities through cross-correlation with standard synthetic template spectra. These velocities are listed in Table 3. Because of their greater precision, the Hectochelle velocities are listed preferentially over those from Hectospec in this table.

For those objects with sufficient $S/N$, we derived integrated velocity dispersions through template cross-correlation as discussed in Strader \etal~(2011). These projected velocity dispersions values ($\sigma_{p}$) are listed in Table 2. The uncertainties in the velocity dispersions are derived from Monte Carlo simulations and include the variance from the use of different templates. 

The use of a finite fiber aperture (for Hectochelle, a radius of $0.75\arcsec = 13.9$ pc) requires a correction from the measured velocity dispersion $\sigma_{p}$ to the global value $\sigma_{\infty}$ used in the virial theorem. At the distance of NGC 4449, this correction is only noticeable for the largest clusters, and was estimated by integrating the appropriate King model over the fiber aperture. The effects of seeing, by contrast, are entirely negligible for these objects.

Given that the core radii are not well-determined for all of our clusters, we restrict our dynamical mass estimates to the virial theorem, which uses the half-light radius ($r_h$) and $\sigma_{\infty}$ (see Strader \etal~2009; 2011). These virial masses are also given in Table 2.

The sample of clusters with dynamical masses includes both young and old clusters. Of the nine objects with measured velocity dispersions, three can be classified as young clusters from their colors, and these clusters are listed separately in Table 2. One of the young clusters (A21) has a published velocity dispersion from Larsen \etal~(2004), where it is called N4449-47. Our estimated dispersion for this cluster ($\sigma_p = 6.6\pm0.5$ \kms) agrees well with the value from Larsen \etal~($\sigma_p = 6.7\pm1.1$ \kms), despite the use of templates with different spectral types. Nonetheless, since our set of templates is optimized for old stellar populations, template mismatch is still a source of systematic uncertainty for the younger clusters in our sample. 

All five old clusters with measured dynamical masses are on the periphery of the galaxy; internal extinction is likely to be minimal, so it is possible to use our F555W magnitudes to calculate cluster luminosities and thus $V$-band mass-to-light ratios. These are listed in Table 2 with the other dynamical parameters and discussed in \S 5.4. For the younger clusters, we list dynamical masses, but not $M/L_V$, owing to the uncertain extinction toward these clusters. 

The dynamical mass estimates confirm the presence of numerous massive clusters. The two most massive clusters are A15 ($\sim 7\times10^6~M_{\odot}$), the young nuclear star cluster of NGC 4449, and B13, an old cluster with a mass of $\sim 3 \times10^6~M_{\odot}$. B13 is discussed further in \S 5.

\section{SDSS Cluster Identification and Classification}

In this section we discuss the objects that we selected for spectroscopic observations from SDSS imaging data. These objects are within 15' of NGC~4449 and were color-selected to be as candidate star clusters. Here we show that some of these objects are GCs that trace an extended stellar halo in NGC~4449.

All of the star clusters within the $HST$/ACS field of view can be clearly recognized as resolved star clusters (though their ages are known only roughly from optical colors). However, candidates from the SDSS imaging are unresolved and thus require additional data to identify them as star clusters in NGC~4449.

Hunter \etal~(2005) report a systemic velocity of 205 \kms\ to NGC 4449, which we adopt.\footnote{We find $v_r = 204\pm2$ \kms\ for A15, the nuclear star cluster of NGC 4449, consistent with the Hunter \etal~value.} This systemic velocity is low enough that objects associated with NGC 4449 cannot be identified by velocity alone, although it does help.  A Besan\c{c}on model of foreground stars with our selection criteria suggests that nearly all foreground stars ($97$\%) will have radial velocities $< 100$\kms. There are 78 SDSS-selected objects that have measurable radial velocities from our Hectospec or Hectochelle data. Of these, 13 have velocities $> 100$ \kms. Two have emission line spectra and blue colors (S59, S115) and are almost certainly associated with star formation in the western edge of NGC 4449.

Two others (S73, S162) have blue colors consistent with those of young star clusters, but are 3--4 kpc north of the galaxy, far outside of its main body, and have radial velocities just above our cut of 100 \kms. S920 ($v_r = 143\pm32$ \kms) is located at a much larger projected galactocentric distance than any other candidate ($> 11$ kpc), and has colors too blue for even a metal-poor GC ($g-r = 0.37$). 
These objects may be either foreground stars or star clusters associated with NGC 4449. The joint colors and radial velocities are unusual
for a foreground star classification, but the spectra are consistent with this interpretation: S73 appears to be an F star of moderate metallicity and S162 a metal-poor halo star (the spectrum of S920 is ambiguous). If any of these objects are indeed star clusters associated with NGC 4449,  their outlying location could be due to their association with tidal material from a recent galactic interaction, and the parameters
of the putative interaction could be constrained by their phase space data.

This analysis leaves 8 candidate old GCs at large radii in NGC 4449 (at these galactocentric distances, the presence of intermediate-age objects would be unexpected). They are relatively homogeneous in their properties: all have $21.0 < g < 19.5$ and radial velocities $\pm 70$ \kms\ of systemic. The mean velocity of these 8 GCs is 206 \kms\ with a dispersion of 40 \kms, consistent with the idea that many of these objects are indeed GCs associated with the galaxy. 

Fortuitously, two of these candidates (S117, S267) are located in archival $HST$/WFPC2 images in the vicinity of NGC 4449 and can be unambiguously classified as GCs that resolve into bright giants. Another, S121, has a a measurable velocity dispersion ($\sigma_p = 5.3\pm0.8$) consistent with that expected for a GC with a typical half-light radius. All three of these GCs are within 20 \kms\ of the systemic velocity of NGC 4449, perhaps suggesting that those candidates with more deviant velocities (e.g., S464 and S186) are less likely to be members, but this is inconclusive because of the small number of objects. Of the remaining five candidates, S125 has a velocity only 20\kms\ from systemic and colors that are nearly identical to those of the confirmed GCs. We classify it as a probable cluster. The other objects have photometry less consistent with the confirmed GCs or a more discrepant radial velocity, so we classify them as possible GCs. The SDSS photometry and these classifications are listed in Table 4.

Spatially, the candidates in Table 4 appear to be elongated along the approximate minor axis of NGC 4449 (see the large red squares in Figure 2). Intriguingly, along this axis, there is a faint diffuse overdensity to the southeast, centered at a radius of 9--10\arcmin. Mart{\'{\i}}nez-Delgado \etal~(2011) show that this feature is a stellar stream. With our current data, any association between outer GCs and this stream is still uncertain.

\begin{figure*}
	\epsscale{1.0}
	\plotone{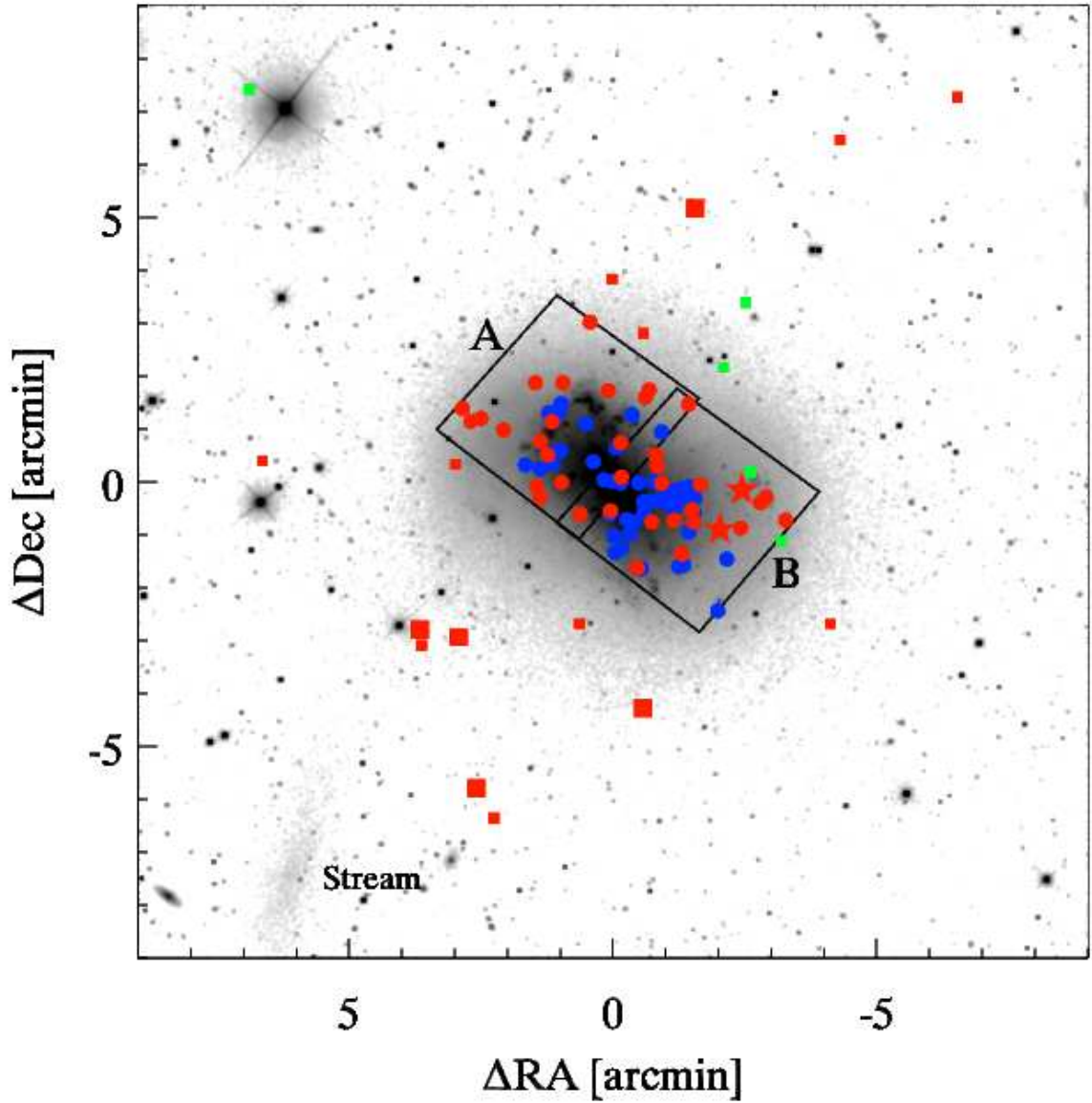}
\figcaption[spat.eps]{\label{fig:fig_s}{
The spatial distribution of clusters discussed in our paper.  Clusters are color coded according to age, with blue objects representing young clusters and red objects representing old ($>$1~Gyr) clusters.  Green objects are the class 4 and 5 objects in Table~4 and represent either young clusters or Galactic foreground objects. Circles represent the good candidate clusters selected from HST data, with B13 and B15 (discussed in \S5) represented as stars.  Large squares represent likely GCs with appropriate radial velocities (class 1 and 2 in Table~4) and the first object in Table~5, while smaller square represent possible clusters from Tables~4 \& 5.  The underlying image is an $r$ band image from the SDSS.  The two HST observations used to select cluster candidates are shown in boxes, while the stream (\S4; Mart{\'{\i}}nez-Delgado \etal~2011) is labeled at the SE corner of the image.}
}
\end{figure*}

There are still a number of objects outside of the $HST$/ACS field of view that have SDSS photometry consistent with being GCs but for which we have not yet obtained spectra. These are listed in Table 5. Serendipitously, one of these objects (SC1) is in an archival WFPC2 image and is a definite GC that resolves into giants in the cluster outskirts. The remaining objects in the table are promising candidates for further observations.

\section{A Detailed Look at Two Massive Clusters}

Of those clusters with $HST$/ACS colors suggesting that they are not young ($\ga 1$ Gyr), the two most luminous objects are B13 and B15. If a substantial group of intermediate-age (1--5 Gyr) clusters were to exist in this galaxy, these luminous clusters might be expected to be part of such a population, since younger stellar populations are brighter for a given stellar mass. In this section, by comparing integrated photometry, spectroscopy, and dynamical mass estimates, we show that these clusters do appear to be authentic old GCs.

One of these two massive clusters, B15, is surrounded by an unusual shell-like distribution of young stars in the $HST$/ACS images. Annibali \etal~(2008) and Annibali \etal~(2011b)\footnote{In these papers the object is called ``Cluster 77".} suggest that it could be the nucleus of a putative dwarf galaxy being tidally disrupted in the galactic potential of NGC 4449. This would be analogous to the nucleus of the Sagittarius dwarf galaxy in the Milky Way (Sarajedini \& Layden 1995), which contains both old metal-poor stars (the globular cluster M54) and a significant population of younger stars with ages ranging from 6 Gyr to $<$1 Gyr (Siegel \etal~2007). However, it is not definitively known whether these younger stars are directly associated with M54 or just with the Sagittarius field; they appear to have the same radial velocity and be spatially coincident with M54 but may be at a different distance (Monaco \etal~2005; Bellazzini \etal~2008; Carretta \etal~2010; Siegel \etal~2011). In any case, a wide range of ages is typical of nuclear star clusters (Seth \etal~2006; Walcher \etal~2006). We keep this possible interpretation in mind when evaluating the observational data below.

\subsection{Low-Resolution Spectroscopy}

A visual comparison of the Hectospec data for B13 and B15 with spectra of M31 clusters (see Caldwell \etal~2011) revealed a remarkable match in detail between the two NGC 4449 GCs and M31 GCs of intermediate metallicity.

We use Lick line indices for a refined spectral analysis. We estimated the spectroscopic ages and metallicities of B13 and B15 using the Hectospec data and the {\tt EZ\_Ages} stellar population analysis program (Graves \& Schiavon 2008). This program uses the single stellar population models of Schiavon (2007). We first smoothed the spectra to the Lick resolution and measured equivalent widths in the passbands defined by Worthey \etal~(1994). These line indices were then converted to the Lick system using the Hectospec zeropoints listed in Schiavon \etal~(2011). The indices are listed in Table 6. {\tt EZ\_Ages} uses the Lick indices in a sequential grid inversion algorithm, comparing index pairs to iteratively solve for the age, [Fe/H], then other abundance ratios (such as [Mg/Fe] and [N/Fe]) that affect fewer indices. The overriding assumption is that there is a single age and abundance pattern for all the stars in the cluster under consideration; the solution returned for a composite population would be a wavelength-dependent light-weighted fit.

For B13 the best-fit age and metallicity are $7.1\pm0.5$ Gyr and [Fe/H] = $-1.12\pm0.06$ dex. The corresponding estimates for B15 are $11.6\pm1.8$ Gyr and [Fe/H] = $-1.12\pm0.06$. These uncertainties do not include systematic errors in the stellar population models, and the presence of hot stars in old clusters (from blue stragglers or blue horizontal branch stars) can lead to the derivation of younger spectroscopic ages (e.g., Schiavon \etal~2004; Cenarro \etal~2008). Thus, it is possible that these two clusters have similar old ages, with B13 having a larger fraction of hot stars. In this case the metallicity estimate for B13 would be biased slightly low (due to the age--metallicity degeneracy), but this is a minor effect.

Of the abundance ratios determined by {\tt EZ\_Ages}, [Mg/Fe] is well-constrained for these two clusters. Both have slightly less than solar [Mg/Fe], with $-0.1 \pm 0.1$ dex for B13 and $-0.2 \pm 0.1$ for B15. The mean [$\alpha$/Fe] ratio for Galactic GCs is $\sim +0.3$; the mean value of [Mg/Fe] is similar but with a larger spread, probably due to abundance variations on the upper red giant branch due to internal mixing processes (Pritzl \etal~2005; see also Colucci \etal~2009 for analysis of a small sample of M31 GCs). Nonetheless, very few clusters have the subsolar [Mg/Fe] ratio found for our two NGC 4449 clusters.

The metallicities may also be  estimated empirically, without reference to stellar population models, by using the relation between two Lick Fe indices and [Fe/H] found for Milky Way GCs in Caldwell \etal~(2011). Using this relation, we derive [Fe/H] = $-1.1\pm0.1$ and $-1.0\pm0.1$ for B13 and B15, respectively, in good agreement with the values from {\tt EZ\_Ages}. This correlation builds in the age--metallicity relation for Milky Way GCs and its use assumes that the NGC 4449 GCs have similar ages to Galactic clusters. Generally, the spectroscopy favors old ages and intermediate metallicities for these two massive clusters.

Four other clusters in our sample have colors suggesting they are old and are sufficiently luminous ($V < 20$) that the Hectospec data have high enough S/N to determine empirical metallicities using this same calibration. The clusters and estimated metallicities are: A6 ($-1.7\pm0.4$), A8 ($-1.2\pm0.2$), A9 ($-1.2\pm0.2$), and B9 ($-0.9\pm0.2$). These [Fe/H] values are similar to those derived for B13 and B15, indicating that intermediate metallicities may be typical for the luminous GCs in NGC 4449.

\subsection{Integrated Photometry}

A complementary method to estimate age and metallicity for clusters is the comparison of panchromatic photometry and appropriate stellar population models. We start with optical data. NGC 4449 is located in the standard imaging footprint of SDSS DR6. Unfortunately, B13 and B15 are too close to the center of the galaxy for proper photometry in the SDSS DR6 photometric catalog. We downloaded the processed SDSS images and and performed aperture photometry on the two clusters with 5-pixel ($1.98\arcsec$) apertures; neither cluster is resolved in the SDSS images. Aperture corrections were determined using bright, isolated stars on the images. The resulting $ugriz$ magnitudes and errors are listed in Table 7.

The combination of UV, optical, and infrared photometry is generally considered to offer additional leverage in breaking the age--metallicity degeneracy for single stellar populations (e.g., de Grijs \etal~2003). These clusters were not sufficiently bright to enable accurate near-IR photometry from 2MASS. However, Spitzer/IRAC maps with the necessary depth exist. Both B13 and B15 are clearly visible in the IRAC channel 1 (3.6 $\mu$m) and channel 2 (4.5 $\mu$m) maps of NGC 4449. We carried out aperture photometry of these sources on calibrated mosaics downloaded from the NASA/IPAC Infrared Science Archive. Because of the presence of nearby sources, a relatively small 4-pixel ($1.2\arcsec$) aperture was used, with aperture corrections taken from Reach \etal~(2005). The IRAC AB magnitudes are listed in Table 7.

Both B13 and B15 lie away from the main body of the galaxy, and have no excess  diffuse emission near their positions in the MIPS $24\mu$m image of the galaxy; we therefore assume that internal reddening is negligible for these clusters.

\subsection{Single Stellar Population Analysis of the Photometry}

We simultaneously fit the SDSS and IRAC photometry of B13 and B15 using the flexible stellar population synthesis (FSPS) models of Conroy \etal~(2009; see also Conroy \& Gunn 2010). These models use Padova isochrones (Marigo \etal~2008) and the BaSeL stellar library; for old ages, the models develop a blue horizontal branch, and we have assumed a fraction of 50\% blue horizontal branch stars for [Fe/H] $\le -1$ and 10\% for higher metallicities. We assumed a Kroupa initial mass function. The photometry was fit in a grid of ages from 100 Myr to 15 Gyr and [Fe/H] from approximately $-2$ to +0.2, and with a softening error of 0.02 mag. Only modest efforts have been made to test standard stellar population models in the IRAC bands, so all of our conclusions are necessarily preliminary.

B13 is consistent with an old, intermediate-metallicity cluster, with an age of $8.5\pm1.5$ Gyr, [Fe/H] = $-0.65\pm0.15$ dex, and a mass of $2.2\pm0.3 \times 10^{6} M_{\odot}$,  comparable to that of $\omega$ Cen. Because of the usual age--metallicity degeneracy, these uncertainties are correlated; an older age would imply a lower metallicity and a higher stellar mass.

The results for B15 are less clear. The photometry appears to be well-fit for two distinct parameter sets. The first is similar to that for B13, old and of intermediate metallicity: a formal age of $10.5\pm2.2$ Gyr; [Fe/H] = $-0.8\pm0.1$ dex. The second is, again consistent with the age--metallicity degeneracy, much younger and more metal-rich: an age of $1.4\pm0.2$ Gyr and [Fe/H] = $-0.4\pm0.2$ dex. The latter is associated with a much lower (but still notable) mass of $2.9\pm0.6 \times10^{5} M_{\odot}$; the old age yields a mass of $1.6\pm0.2 \times10^{6} M_{\odot}$. For both clusters, the models provide good fits, with rms differences of $< 3$\% for B13 and about 4\% for both intermediate-age and old models for B15. These fits are shown in Figure 3. Of course, as with all stellar population modeling, the quality of the fits do not indicate that the best-fit parameters are necessarily accurate. We also note that in both cases the photometric metallicities are higher than those derived from spectroscopy. There are many possible sources for this systematic difference, such as the assumed properties of the hot stars in the stellar population models.

If indeed B15 is the nucleus of a dwarf galaxy being accreted by NGC 4449, the presence of an extended star formation history would not be unexpected.

\begin{figure}
  \begin{center}	
	\epsscale{1.2}
	\plotone{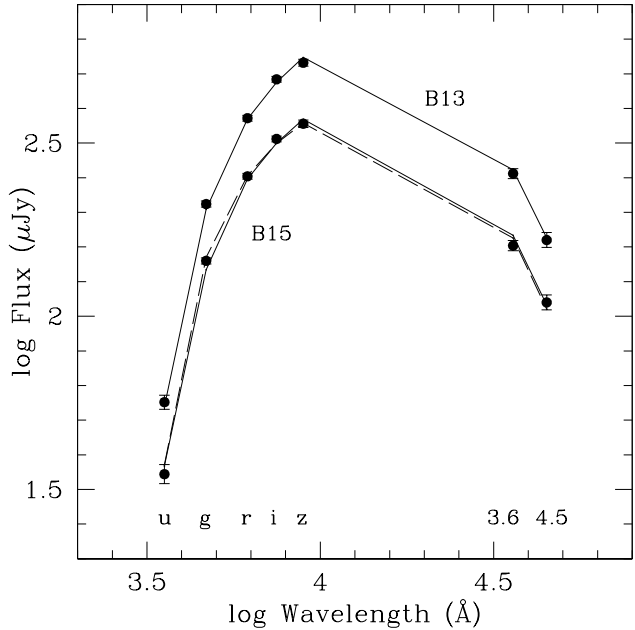}
 	\end{center}
\figcaption[figx]{\label{fig:fig_x}
FSPS model fits (lines) to measured SDSS+IRAC magnitudes for B13 and B15 (filled circles), as described in detail in the text. B13 has a single old model plotted with an age of 8.5 Gyr and [Fe/H] = $-0.65$. B15 has two models plotted: the solid line is an old model (10.5 Gyr, [Fe/H] = $-0.8$); the dashed line is a young model (1.4 Gyr, [Fe/H] = $-0.4$). Either model provides a reasonable formal fit.}
\end{figure}

\subsection{Mass-to-Light Ratios}

As discussed in \S 3.2, we use the F555W photometry and Hectochelle dynamical masses to calculate $M/L_V$ for the star clusters that are not young. For B13 and B15, we find $M/L_V$ = $2.83\pm0.53$ and $1.14\pm0.20$, respectively. These values can be readily compared to those observed for old massive GCs in M31 (Strader \etal~2011), as we do in Figure 4. Currently, M31 has the largest sample of GCs with high-quality dynamical masses. The $M/L$ value for B13 is slightly high but generally consistent with that of very massive, intermediate-metallicity M31 GCs. The $M/L_V$ for B15 is typical of that for M31 GCs of its mass and metallicity.

As discussed in Strader \etal~(2009), there are few intermediate-age star clusters with empirical $M/L$ estimates, and so it is difficult to predict what the $M/L_V$ for such clusters would be. Stellar population model predictions assuming a Kroupa IMF for intermediate-metallicity objects suggest that $M/L_V$ would be lower by a factor of 2--4 for intermediate-age clusters than for old clusters (Maraston 2005). In particular, for the ``young" stellar population solution from the photometry for B15, the FSPS prediction is  $M/L_V = 0.40$. Despite the uncertainty in stellar population models, the measured $M/L_V$ values make it unlikely that either B13 or B15 are of intermediate age ($\la 5$~Gyr).

The $M/L_V$ estimates for the other candidates GCs listed in Table 2 (A8, A9, and A22) are in the range $\sim 1.2-1.5$, also consistent with old M31 GCs of comparable mass and metallicity. There is no clear evidence for clusters of intermediate age from this small sample of $M/L_V$ measurements.

\begin{figure}
  \begin{center}	
	\epsscale{1.2}
	\plotone{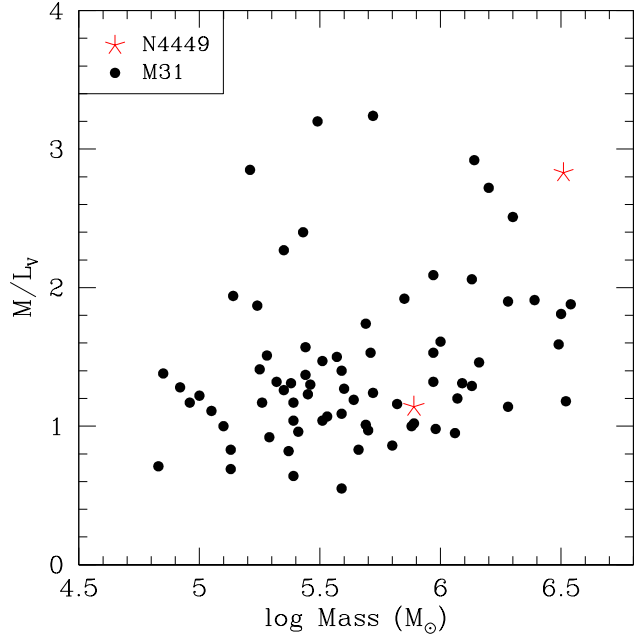}
 	\end{center}
\figcaption[fig2]{\label{fig:fig_2}
$M/L_V$ vs.~dynamical mass for GCs in M31 (filled circles) and for B13 and B15 in NGC 4449 (red stars), adapted from Strader \etal~(2011). We only plot those GCs with [Fe/H] $> -1$ in M31; the $M/L_V$ values for more metal-poor GCs are slightly higher, but with similar mass trends. The decreasing $M/L_V$ toward lower masses is due to dynamical evolution. $M/L_V$ for B13, the more massive cluster, is somewhat high but within the envelope of the M31 observations. The $M/L_V$ estimate for B15, the less massive cluster, is consistent with M31 GCs of similar mass. There is no evidence that either object has the low $M/L$ expected of an intermediate-age cluster.}
\end{figure}

\subsection{Resolved Color-Magnitude Diagrams}

Both B13 and B15 are resolved into giants in their outskirts. To study these stars, we ran the DOLPHOT package (Dolphin 2000) on the $HST$/ACS data to obtain resolved stellar photometry for the clusters and the surrounding area. Due to crowding, the inner regions of the clusters are unsuitable for accurate photometry. Therefore, we examined stars in annuli in the outermost parts of the two clusters. The inner boundaries of these annuli were determined using the sharpness and crowding parameters from DOLPHOT. In B13, almost no stars were detected in the inner $2\arcsec$, with enhanced crowding and blended stars observed to a radius of $\sim3.5\arcsec$.  We adopted this value as our inner radius. The outer radius was set using an integrated surface brightness profile measured with the {\tt ellipse} task in IRAF; the light from B13 was found to merge with the background at a radius of $\sim7\arcsec$, and we use an outer radius of $6.5\arcsec$. Performing a similar procedure for B15 led to inner and outer radii of $2.5\arcsec$ and $4\arcsec$, respectively.

Stars in annuli around each cluster are shown in Figure 5. For both clusters, these outer stars are consistent with being old, modestly metal-poor red giant branch (RGB) stars with [Fe/H] between $-1.3$ and $-0.4$, consistent with the metallicity estimates in \S5.1 and \S5.3. The two clusters look quite similar to each other and the background field stars, with B15 perhaps slightly more metal-poor than the background stars. Because we are studying stars in the outer parts of both clusters, we expect background contamination to be significant, and thus do not attempt quantitative stellar population analysis on the color-magnitude diagrams. 

One notable difference between B13 and B15 in Figure 5 is that B15 has only a single asymptotic giant branch (AGB) star brighter than the tip of the RGB, while B13 has 18 candidate AGB stars (defined as those brighter than the RGB tip and with F555W--F814W $ >1.5$), compared to only 8 in an equal area background annulus centered on the cluster. Using Poisson statistics this is formally a $\sim 2.3\sigma$ excess and marginal evidence for a true enhancement of AGB stars in B13. This overdensity does not depend sensitively on the exact boundaries of the cluster or background annulus, and the spatial distribution of AGB stars is suggestive of a slight enhancement around B13.  If this enhancement is real, it might suggest an intermediate-age stellar population in the cluster (note that, given the mass of B13, it could well have multiple stellar populations). However, due to the location of B13---near a strong stellar density gradient at the edge of the galaxy disk---we do not consider this conclusion secure.

\begin{figure*}
  \begin{center}	
	\epsscale{1.1}
	\plotone{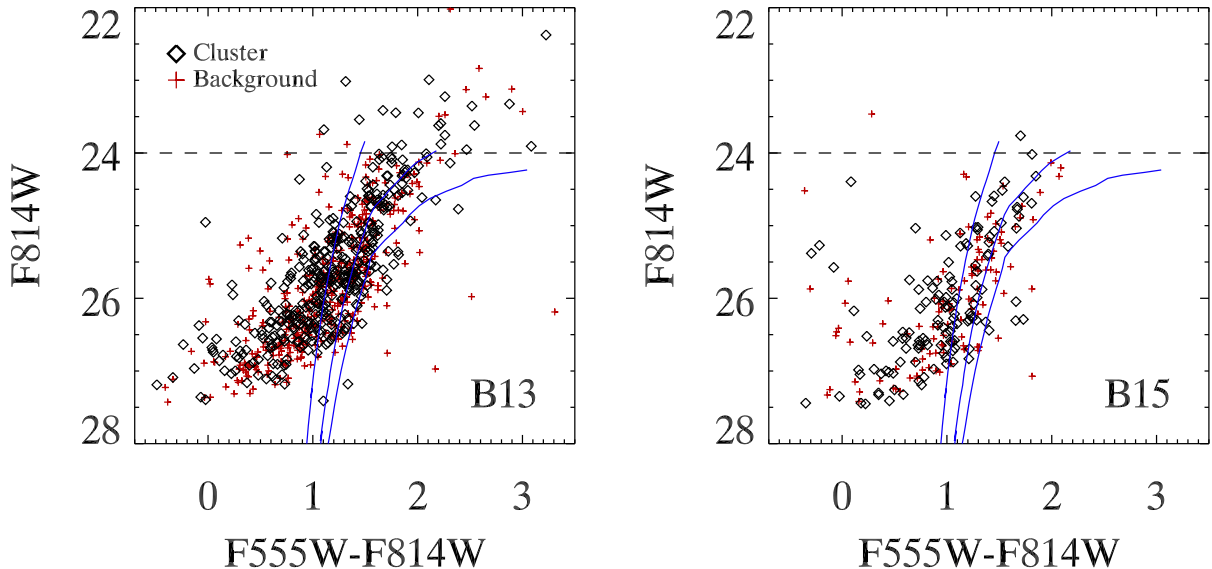}
 	\end{center}
\figcaption[fig3]{\label{fig:fig_3}
F814W vs.~F555W--F814W color-magnitude diagrams of the uncrowded regions around the B13 and B15 clusters. An elliptical aperture with axis ratio of 0.76 and major axis radius between $3.5\arcsec$ and $6.5\arcsec$ ($\sim 65-120$ pc) was used for B13, while for B15 an axis ratio of 0.8 and aperture of $2.5\arcsec$ to $4\arcsec$  ($\sim 46-74$ pc) was used. Stars from a background area at larger radii, matched to the area of the cluster annulus, are shown as crosses. Overplotted are Padova tracks with [Fe/H]=$-1.3$, $-0.7$ and $-0.4$ (Marigo \etal~2008). The luminosity of the tip of the red giant branch is given by the dashed line; all old stars
brighter than this should be AGB stars.}
\end{figure*}

After the submission of the current manuscript, a paper was published with an analysis of the resolved stellar populations in the outskirts of B15 and its surrounding younger candidate ``tidal tails" (Annibali \etal~2011b). They conclude that B15 has an age $\ga 1$ Gyr, consistent with our findings. 

\subsection{Conclusions}

Regarding B13, the low-resolution spectroscopy and photometry both favor an age slightly lower than typical (7--9 Gyr) for an ancient GC. There is also
marginal evidence for an enhancement in AGB stars in the outer regions of the cluster. However, such an enhancement would favor a much younger
age ($\sim$ a few Gyr) than inferred from the other data. In addition, the 7--9 Gyr age could easily be caused by a relatively small population of hot stars in an old (12--13 Gyr) stellar population. B13 is very massive (comparable to $\omega$ Centauri), with an estimated dynamical mass of $\sim 3\times10^{6} M_{\odot}$.

The photometry for B15 is consistent with two solutions, of which one is intermediate-age. This solution is not supported by the spectroscopy and dynamical mass-to-light ratio of the cluster, and it is more likely that the cluster is old and of intermediate metallicity.

As noted earlier, Annibali \etal~(2008) and Annibali \etal~(2011b) suggested that B15 could be the nucleus of a disrupting galaxy.  We find no strong evidence for young or multiple populations in either this cluster or B13.  

For both clusters the spectroscopy favors a slightly lower (by 0.3--0.4 dex) metallicity than the photometry; given the uncertain calibration of the Spitzer/IRAC photometry and larger effect of hot stars on optical colors, we consider the spectroscopic estimates to be more accurate.

Overall, most data are consistent with old ages and intermediate metallicities for both B13 and B15. The limited evidence available for other luminous clusters in NGC 4449 suggests similar stellar populations to these two brightest GCs. 

\section{Discussion}

\subsection{The Overabundance of Massive Star Clusters in NGC 4449}

The LMC is the dwarf galaxy whose star cluster system is perhaps the best studied, so it makes a useful counterpoint to NGC 4449. First we compare the stellar masses of NGC 4449 and the LMC. The extinction-corrected integrated absolute magnitude of NGC 4449 is $M_B = -18.9$ and $M_K = -20.7$ (Paturel \etal~2003; Jarrett \etal~2003). Using mass-to-light ratios from Bell \etal~(2003) for $(B-V)_{0} = 0.33$, the average of the $B$ and $K$-band implied total stellar mass is $2.6\times10^{9} M_{\odot}$. In a review of the properties of the LMC, van der Marel (2004) reports a stellar mass of $2.7\times10^{9} M_{\odot}$, nearly identical to that of NGC 4449. So, despite the obvious differences in recent star formation between NGC 4449 and the LMC, the similarity in their overall stellar masses suggests that a direct comparison of their old star cluster populations is reasonable.S

The LMC has sixteen old star clusters (GCs) with ages $> 5$ Gyr (Mackey \& Gilmore 2004). Fifteen of these have very old ages ($\sim 12$ Gyr); the other is the unusual low-mass cluster ESO121-SC03, which is slightly younger at 9 Gyr (Mackey \etal~2006). Perhaps surprisingly, the integrated luminosities of most of the LMC GCs are known only imprecisely. Here we preferentially use the $V$-band luminosities from McLaughlin \& van der Marel (2005), which are derived from King model fits to the surface brightness profiles of the clusters. Luminosities are available from this source for twelve of the sixteen GCs, including all of the more massive GCs. For the other four GCs we take values compiled in other literature sources (Mackey \& Gilmore 2004; Pessev \etal~2008); these values are less certain.

As discussed in \S 2.1, at the distance of NGC 4449, $V = 20$ corresponds very closely to $M_V = -8$, which we take as a convenient limit for massive GCs. There are five LMC GCs at or above this luminosity; the most luminous is NGC 1916 ($M_V = -8.9$ or  $\sim 6\times10^{5} M_{\odot}$). We do not yet have a full census of GCs in NGC 4449, but we have found at least 19 objects (15 from ACS, 4 from SDSS) above this luminosity for which all available evidence suggests that they are old GCs, with a handful of additional luminous candidates.  Assuming a GC luminosity function similar to that in the Milky Way, these bright clusters would suggest the total number of GCs in NGC~4449 is $\ga 69$.  This corresponds to lower limits on the GC specific frequency (a luminosity-normalized count of GCs) of $S_N \ga 1.5$ and $T \ga 25$, where $T$ is the number of GCs per 10$^9$~M$_\odot$ of galaxy stellar mass.  These GC frequencies are higher than typical late-type spirals (Chandar \etal~2001; 2004), but are not atypical of somewhat fainter dwarf irregular (Seth \etal~2004; Georgiev \etal~2010) and elliptical galaxies (see, e.g., Durrell \etal~1996; Strader \etal~2006; Miller \& Lotz 2007; Peng \etal~2008). The luminosity functions of the old NGC~4449 and LMC GCs are compared in Figure~6.

\begin{figure}
	\epsscale{1.2}
	\plotone{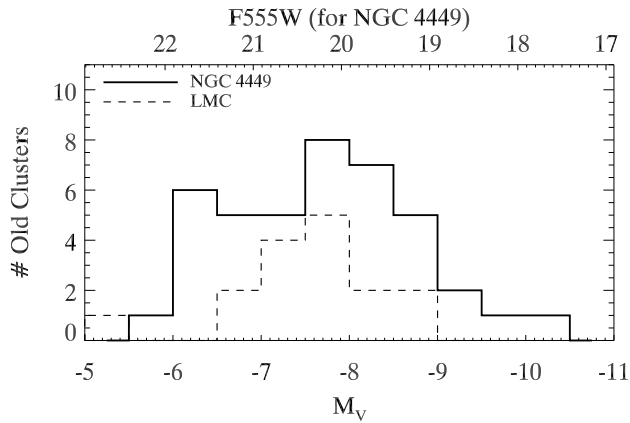}
\figcaption[lumfunc.eps]{\label{fig:fig_s}
The luminosity function of old (ages $\ga 1$~Gyr) star clusters in NGC~4449 (solid line) compared to 16 old ($> 3$ Gyr) GCs in the LMC (dotted line). The solid line shows the luminosity function of good quality old NGC~4449 clusters selected from the HST data (Table~1) plus the confirmed SDSS sources from Table~4 (where $V$ magnitudes have been estimated from SDSS photometry).  The $HST$ sample is complete at $F555W \lesssim 20$, however, it covers only the inner part of the galaxy, and we have included only SDSS objects that have been confirmed via spectroscopy or $HST$ imaging. Even with this incomplete sample, NGC~4449 shows a clear overabundance of massive, old star clusters with respect to the LMC. }

\end{figure}

We cannot rule out the possibility that a subset of the NGC~4449 GCs with $V<20$, or some of the less luminous clusters, are in fact  of intermediate age ($\sim$ 1--5 Gyr). As we state above, depending on their exact age, such objects could have masses 2--4 times lower than old clusters of the same luminosity, and the existence of intermediate-age clusters with masses $\la 10^5 M_{\odot}$ would not be unexpected. Through multi-band photometry (from the UV to the mid-IR), low-resolution spectroscopic age estimates, mass-to-light ratio measurements, and color-magnitude diagrams, we have shown that there is no compelling evidence that either of the two most luminous star clusters in NGC 4449 (B13 and B15) is of intermediate age, though the case of B15 is less certain. If NGC~4449 did possess luminous intermediate-age GCs, then it would be appropriate to consider intermediate-age LMC clusters in a comparison of their cluster systems, and the abundance of the NGC~4449 system would be less remarkable.

The metallicities and chemical abundance ratios of clusters in the LMC and NGC 4449 offer another puzzling comparison. The median metallicity of the 15 old GCs in the LMC is [Fe/H] $\sim -1.9$ (Mackey \& Gilmore 2004). This is much lower than the median spectroscopic metallicity of [Fe/H] $\sim -1.1$ to $-1.2$ for GCs in NGC 4449, even considering that we do not have metallicities for some of the more distant, possibly metal-poor GCs in NGC 4449. If these GCs are generally ancient then the high mean metallicity is extremely surprising.

Similarly, [Mg/Fe] for LMC GCs is typically [$\alpha$/Fe] $\sim 0.2$--0.4 (Johnson \etal~2006), similar to Galactic GCs, compared to the subsolar values observed for B13 and B15 in this paper. The higher values of [Fe/H] and lower values of [Mg/Fe] are evocative of Pal 12, a Galactic GC thought to have been accreted from the Sgr dwarf (Cohen 2004). Pal 12 has an age of $\sim 8$--9 Gyr (Mar{\'{\i}}n-Franch \etal~2009). Therefore, a possible interpretation is that many of the GCs in NGC 4449 could have been made in a burst of cluster formation a few Gyr later than the old (12--13 Gyr) metal-poor GCs observed in both the Milky Way and the LMC.

\subsection{Stellar Halos and Metal-poor Globular Clusters}

The overwhelming weight of evidence now supports the idea that the metal-poor stellar halo of the Milky Way was built largely through the accretion of less massive galaxies, partially via the direct contribution of stars from these dwarf galaxies, and partially from \emph{in situ} stars propelled to larger radii by the accretion events (Searle \& Zinn 1978; Ibata \etal~1994; Bullock \& Johnston 2005; Zolotov \etal~2009; Carollo \etal~2010). Similar accretion events are observed in M31 (Ibata \etal~2001) and other nearby disk galaxies (Mart{\'{\i}}nez-Delgado \etal~2010). The metal-poor GC systems of these galaxies trace their stellar halo stars in both
chemistry and spatial distribution (Brodie \& Strader 2006).

Many dwarf galaxies also have extended ``halos" of stars, if one considers stellar components that extend above the outer extrapolation of the inner surface brightness profile, but the relationship between these halos and those observed in more massive galaxies is unclear. Such a halo has indeed been observed in NGC 4449 (Ry{\'s} \etal~2011). Theoretically, accretion is expected to be less effective in producing stellar halos around dwarf galaxies because of the decreasing efficiency of star formation in less massive halos (Purcell \etal~2007). Alternative methods for the formation
of stellar halos in dwarf galaxies, such as by internal feedback, have been proposed (e.g., Stinson \etal~2009).

Ry{\'s} \etal~(2011) do not find evidence for a stellar population gradient in the halo of NGC 4449, with no change in the color of the RGB or the relative number of carbon stars. Thus there is no direct evidence for a metal-poor component in this halo. However, the existence of metal-poor GCs shows that such stars are in fact present. Using the color--metallicity relation of Sinnott \etal~(2010), the $g-z$ colors of the four confirmed outer SDSS GCs correspond to metallicities of [Fe/H] = $-1.1$ to $-1.6$. These findings add to a growing picture that dwarf galaxies (e.g., NGC 6822; Hwang \etal~2011) have halo GCs.

As mentioned above, there is separate evidence for ongoing accretion (a luminous stellar stream) in NGC 4449 (Mart{\'{\i}}nez-Delgado \etal~2011). The combination of these two observations---a metal-poor stellar population and a stellar steam---are consistent with the idea that the halo of NGC 4449 could indeed have been built by accretion, just as is the case for more massive galaxies. The discovery of more distant GCs (perhaps requiring $HST$) should enable kinematic studies of the halo GC system of the galaxy, which may help distinguish between the \emph{in situ} and accretion models for the formation of stellar halos in dwarf galaxies.

\acknowledgments

We thank David Sand for help with data acquisition. We thank Susan Tokarz for help with the spectroscopic data reduction, and Perry Berlind and Michael Calkins for help acquiring the data. Ricardo Schiavon helped us to operate his EZ\_Ages program. Some of the observations reported here were obtained at the MMT Observatory, a joint facility of the Smithsonian Institution and the University of Arizona. This paper uses data products produced by the OIR Telescope Data Center, supported by the Smithsonian Astrophysical Observatory. Based on observations made with the NASA/ESA Hubble Space Telescope, and obtained from the Hubble Legacy Archive, which is a collaboration between the Space Telescope Science Institute (STScI/NASA), the Space Telescope European Coordinating Facility (ST-ECF/ESA) and the Canadian Astronomy Data Centre (CADC/NRC/CSA). This work is based in part on observations made with the Spitzer Space Telescope, which is operated by the Jet Propulsion Laboratory, California Institute of Technology under a contract with NASA. This research has made use of the NASA/IPAC Infrared Science Archive, which is operated by the Jet Propulsion Laboratory, California Institute of Technology, under contract with the NASA. We thank the anonymous referee for a helpful report.

\LongTables
\newpage

\begin{deluxetable}{lccccccccccc}
\tablewidth{0pt}
\tabletypesize{\tiny}
\setlength{\tabcolsep}{0.005in} 
\tablecaption{HST/ACS Star Cluster Photometry
        \label{tab:phot_acs}}
\tablehead{ID & R.A.~(J2000) & Dec.~(J2000) & $r_{ap}$\tablenotemark{a} & $r_h$\tablenotemark{b} & F555W & F435W--F555W & F555W--F814W & Age\tablenotemark{c} & G01\tablenotemark{d} & A11\tablenotemark{e} & R11\tablenotemark{f} \\
                          &  (deg)               & (deg)    & (\arcsec) & (\arcsec)   & (mag) &  (mag) & (mag) & & & & \\}
 
\startdata
A0  & 187.112493 & 44.116711 & 1.10 & 0.30 &  $20.26\pm0.03$  & 0.69 & 1.05 &  Old  & \nodata & 1 & \nodata \\
A1  & 187.108833 & 44.112647 & 0.89 & 0.38 &  $21.51\pm0.03$  & 0.71 & 1.01 &  Old  & \nodata & \nodata & \nodata \\
A2  & 187.104277 & 44.113544 & 1.11 & 0.34 &  $21.40\pm0.03$  & 0.62 & 0.93 &  Old  & \nodata & \nodata & \nodata \\
A3  & 187.084846 & 44.098949 & 0.74 & 0.42 &  $21.89\pm0.07$  & 0.36 & 0.71 &  Young  & \nodata & \nodata & \nodata \\
A4  & 187.078377 & 44.097657 & 0.70 & 0.34 &  $20.93\pm0.03$  & 0.13 & 0.35 &  Young  & \nodata & \nodata & \nodata \\
A5  & 187.079791 & 44.092120 & 0.89 & 0.31 &  $20.66\pm0.03$  & 0.80 & 1.17 &  Old  & 61 & 19 & \nodata \\
A6  & 187.078246 & 44.088684 & 1.24 & 0.44 &  $19.66\pm0.03$  & 0.71 & 1.05 &  Old  & 60 & 20 & 8 \\
A7  & 187.068940 & 44.093407 & 0.74 & 0.22 &  $19.64\pm0.04$  & 0.71 & 1.00 &  Old  & 53 & 21 & 18 \\
A8  & 187.060675 & 44.083354 & 0.58 & 0.18 &  $19.37\pm0.03$  & 0.72 & 1.07 &  Old  & 48 & 39 & 3 \\
A9  & 187.078096 & 44.106327 & 0.91 & 0.23 &  $19.40\pm0.04$  & 0.71 & 1.04 &  Old  & 59 & 8 & 64 \\
A10  & 187.073031 & 44.112500 & 0.86 & 0.44 &  $19.98\pm0.07$  & 0.52 & 0.98 &  Old  & \nodata & 7 & 74 \\
A11  & 187.072770 & 44.099288 & 0.78 & 0.24 &  $19.54\pm0.04$  & 0.12 & 0.39 &  Young  & 57 & 17 & 51 \\
A12  & 187.069359 & 44.103300 & 0.73 & 0.24 &  $20.01\pm0.03$  & 0.15 & 0.43 &  Young  & 54 & 15 & 59 \\
A13  & 187.072929 & 44.102797 & 0.76 & 0.47 &  $20.16\pm0.17$  & 0.16 & 0.45 &  Young  & 58 & 11 & \nodata \\
A14  & 187.054929 & 44.099943 & 0.67 & 0.35 &  $18.18\pm0.03$  & 0.04 & 0.26 &  Young  & \nodata & 26 & 52 \\
A15  & 187.046198 & 44.093642 & 0.70 & 0.28 &  $15.46\pm0.11$  & 0.36 & 0.88 &  Young  & 1 & SSC & 21 \\
A16  & 187.042936 & 44.093183 & 0.51 & 0.21 &  $18.38\pm0.03$  & 0.33 & 0.75 &  Young  & 27 & 48 & 17 \\
A17  & 187.042518 & 44.095060 & 0.68 & 0.17 &  $19.99\pm0.07$  & 0.63 & 1.02 &  Old  & \nodata & 46 & 22 \\
A18  & 187.080292 & 44.124823 & 0.68 & 0.24 &  $21.47\pm0.03$  & 0.65 & 0.95 &  Old  & \nodata & 2 & \nodata \\
A19  & 187.068282 & 44.124788 & 1.44 & 0.45 &  $20.13\pm0.03$  & 0.71 & 1.06 &  Old  & \nodata & 3 & \nodata \\
A20  & 187.070128 & 44.115388 & 1.42 & 0.50 &  $18.18\pm0.03$  & 0.20 & 0.46 &  Young  & \nodata & 6 & 79 \\
A21  & 187.058126 & 44.111920 & 1.30 & 0.26 &  $17.54\pm0.03$  & 0.23 & 0.55 &  Young  & 47 & 18 & 72 \\
A22  & 187.042842 & 44.105827 & 0.74 & 0.25 &  $18.52\pm0.03$  & 0.71 & 1.04 &  Old  & 26 & 34 & 63 \\
A23  & 187.048364 & 44.122332 & 1.05 & 0.19 &  $19.65\pm0.03$  & 0.64 & 0.95 &  Old  & \nodata & 14 & 88 \\
A24  & 187.037870 & 44.114695 & 1.10 & 0.39 &  $17.59\pm0.03$  & 0.14 & 0.15 &  Young  & \nodata & \nodata & \nodata \\
A25  & 187.056230 & 44.143925 & 1.25 & 0.53 &  $20.60\pm0.04$  & 0.82 & 0.96 &  Old  & \nodata & \nodata & \nodata \\
A26  & 187.031991 & 44.120298 & 1.15 & 0.21 &  $19.67\pm0.03$  & 0.74 & 1.05 &  Old  & \nodata & 27 & 86 \\
A27  & 187.030319 & 44.122556 & 0.92 & 0.30 &  $20.39\pm0.03$  & 0.75 & 1.04 &  Old  & \nodata & 24 & 89 \\
A28  & 187.013005 & 44.118046 & 0.78 & 0.41 &  $21.51\pm0.03$  & 0.79 & 1.09 &  Old  & \nodata & \nodata & \nodata \\
A29  & 187.094373 & 44.109993 & 1.75 & 0.75 &  $21.52\pm0.17$  & 0.69 & 1.03 &  Old  & \nodata & \nodata & \nodata \\
A33  & 187.074361 & 44.115292 & 0.61 & 0.35 &  $20.61\pm0.04$  & 0.33 & 0.70 &  Young  & \nodata & 4 & 78 \\
A34  & 187.074916 & 44.102104 & 0.67 & 0.49 &  $21.10\pm0.10$  & 0.59 & 1.21 &  Old  & \nodata & \nodata & \nodata \\
A43  & 187.069080 & 44.118139 & 0.72 & 0.16 &  $20.91\pm0.05$  & 0.26 & 0.55 &  Young  & \nodata & 5 & 82 \\
A45  & 187.044746 & 44.104210 & 0.59 & 0.27 &  $20.56\pm0.03$  & 0.19 & 0.48 &  Young  & \nodata & 33 & \nodata \\
A46  & 187.024903 & 44.109298 & 0.92 & 0.23 &  $22.69\pm0.09$  & 0.32 & 0.49 &  Young  & \nodata & \nodata & \nodata \\
B1  & 187.038746 & 44.077399 & 1.06 & 0.35 &  $19.05\pm0.05$  & 0.23 & 0.50 &  Young  & \nodata & 67 & 95 \\
B2  & 187.035500 & 44.066435 & 0.93 & 0.58 &  $20.67\pm0.05$  & 0.54 & 1.11 &  Old  & \nodata & \nodata & \nodata \\
B3  & 187.050271 & 44.094154 & 0.60 & 0.30 &  $17.77\pm0.03$  & 0.26 & 0.94 &  Young  & 37 & 38 & 26 \\
B4  & 187.036566 & 44.081005 & 0.72 & 0.35 &  $19.62\pm0.03$  & 0.17 & 0.41 &  Young  & 19 & 65 & 99 \\
B5  & 187.040380 & 44.081728 & 0.74 & 0.16 &  $19.68\pm0.21$  & 0.25 & 0.60 &  Young  & \nodata & \nodata & 126 \\
B6  & 187.032557 & 44.084227 & 0.81 & 0.43 &  $19.37\pm0.03$  & 0.20 & 0.54 &  Young  & \nodata & 66 & 124 \\
B7  & 187.029069 & 44.080953 & 0.76 & 0.22 &  $21.00\pm0.05$  & 0.72 & 1.10 &  Old  & \nodata & 69 & \nodata \\
B8  & 187.019341 & 44.081410 & 0.85 & 0.19 &  $20.19\pm0.03$  & 0.80 & 1.10 &  Old  & 8 & 72 & 100 \\
B9  & 187.015970 & 44.070817 & 1.17 & 0.32 &  $19.35\pm0.03$  & 0.72 & 1.06 &  Old  & 7 & 76 & 92 \\
B10  & 187.026655 & 44.098535 & 0.63 & 0.22 &  $18.73\pm0.03$  & 0.72 & 1.06 &  Old  & 13 & 58 & 121 \\
B11  & 187.030670 & 44.087416 & 0.39 & 0.22 &  $20.83\pm0.03$  & 0.29 & 0.57 &  Young  & 17 & 63 & 108 \\
B12  & 187.010694 & 44.080877 & 0.83 & 0.30 &  $20.57\pm0.03$  & 0.64 & 0.98 &  Old  & 5 & 75 & 128 \\
B13  & 186.999279 & 44.078597 & 2.80 & 0.39 &  $17.61\pm0.03$  & 0.72 & 1.04 &  Old  & \nodata & 79 & 96 \\
B14  & 186.990082 & 44.079038 & 1.03 & 0.26 &  $19.79\pm0.03$  & 0.68 & 1.02 &  Old  & \nodata & 80 & 97 \\
B15  & 186.989519 & 44.091054 & 1.67 & 0.25 &  $18.16\pm0.03$  & 0.70 & 1.02 &  Old  & \nodata & 77 & 115 \\
B16  & 186.970308 & 44.081423 & 0.91 & 0.18 &  $20.26\pm0.03$  & 0.60 & 0.93 &  Old  & \nodata & 81 & 101 \\
B18  & 186.978912 & 44.088765 & 0.69 & 0.36 &  $21.60\pm0.03$  & 0.64 & 0.97 &  Old  & \nodata & \nodata & \nodata \\
B19  & 187.027480 & 44.102077 & 0.72 & 0.19 &  $19.06\pm0.03$  & 0.74 & 1.06 &  Old  & 14 & 52 & 122 \\
B20  & 187.047356 & 44.084428 & 0.75 & 0.17 &  $21.05\pm0.13$  & 0.86 & 1.29 &  Old  & \nodata & \nodata & \nodata \\
B21  & 187.047568 & 44.083394 & 0.55 & 0.24 &  $19.80\pm0.05$  & 0.13 & 0.38 &  Young  & 34 & 53 & 102 \\
B25  & 187.042418 & 44.072863 & 0.85 & 0.23 &  $20.33\pm0.10$  & 0.20 & 0.03 &  Young  & \nodata & \nodata & \nodata \\
B26  & 187.045392 & 44.071233 & 0.63 & 0.46 &  $22.12\pm0.06$  & 0.10 & 0.31 &  Young  & \nodata & \nodata & \nodata \\
B27  & 187.033747 & 44.066297 & 0.79 & 0.56 &  $21.53\pm0.10$  & 0.12 & 0.31 &  Young  & \nodata & \nodata & \nodata \\
B28  & 187.045953 & 44.076556 & 0.56 & 0.23 &  $21.35\pm0.15$  & 0.02 & 0.34 &  Young  & \nodata & \nodata & \nodata \\
B29  & 187.017293 & 44.066895 & 0.60 & 0.30 &  $22.37\pm0.12$  & 0.05 & 0.19 &  Young  & \nodata & \nodata & \nodata \\
B30  & 187.015244 & 44.067279 & 0.65 & 0.29 &  $22.25\pm0.03$  & 0.21 & 0.43 &  Young  & \nodata & \nodata & \nodata \\
B31  & 187.000105 & 44.052952 & 0.46 & 0.20 &  $22.90\pm0.03$  & 0.48 & 0.77 &  Young  & \nodata & \nodata & \nodata \\
B32  & 187.034906 & 44.093247 & 0.64 & 0.19 &  $20.93\pm0.12$  & 0.33 & 0.74 &  Young  & \nodata & \nodata & 118 \\
B33  & 187.025826 & 44.087674 & 0.61 & 0.21 &  $20.94\pm0.04$  & 0.16 & 0.41 &  Young  & 11 & 68 & 109 \\
B34  & 187.023023 & 44.088887 & 0.60 & 0.22 &  $21.58\pm0.05$  & 0.17 & 0.44 &  Young  & 9 & \nodata & \nodata \\
B35  & 187.021171 & 44.087908 & 0.68 & 0.24 &  $21.62\pm0.05$  & 0.23 & 0.42 &  Young  & \nodata & \nodata & \nodata \\
B36  & 187.019344 & 44.085377 & 0.59 & 0.22 &  $21.89\pm0.08$  & 0.10 & 0.35 &  Young  & \nodata & \nodata & \nodata \\
B37  & 187.021277 & 44.086101 & 0.52 & 0.24 &  $22.01\pm0.06$  & 0.05 & 0.34 &  Young  & \nodata & \nodata & \nodata \\
B38  & 187.028162 & 44.093595 & 0.46 & 0.27 &  $22.53\pm0.13$  & 0.49 & 0.66 &  Young  & \nodata & \nodata & \nodata \\
B39  & 187.032845 & 44.087273 & 0.76 & 0.38 &  $21.19\pm0.13$  & 0.29 & 0.73 &  Young  & \nodata & \nodata & \nodata \\
B41  & 187.012944 & 44.077859 & 0.53 & 0.26 &  $22.20\pm0.04$  & 0.16 & 0.50 &  Young  & \nodata & \nodata & \nodata \\
B43  & 187.010671 & 44.083738 & 0.78 & 0.69 &  $22.46\pm0.16$  & 0.33 & 0.39 &  Young  & \nodata & \nodata & \nodata \\
B44  & 186.996171 & 44.069247 & 0.45 & 0.10 &  $21.80\pm0.03$  & 0.39 & 0.49 &  Young  & \nodata & \nodata & \nodata \\
B46  & 187.007939 & 44.092770 & 1.18 & 0.63 &  $21.11\pm0.03$  & 0.70 & 1.02 &  Old  & \nodata & \nodata & \nodata \\
B47  & 187.012683 & 44.092107 & 1.27 & 0.54 &  $19.64\pm0.03$  & 0.18 & 0.38 &  Young  & \nodata & 71 & 117 \\
B48  & 187.010639 & 44.090803 & 0.85 & 0.33 &  $21.03\pm0.05$  & 0.29 & 0.74 &  Young  & \nodata & \nodata & \nodata \\
B49  & 187.010370 & 44.088370 & 0.91 & 0.33 &  $21.33\pm0.07$  & 0.09 & 0.22 &  Young  & 3 & 74 & \nodata \\
B50  & 187.019873 & 44.090487 & 0.69 & 0.32 &  $21.36\pm0.05$  & 0.30 & 0.72 &  Young  & \nodata & \nodata & \nodata \\
B51  & 187.016756 & 44.091377 & 0.61 & 0.22 &  $22.29\pm0.24$  & 0.04 & 0.16 &  Young  & \nodata & \nodata & \nodata \\
B52  & 187.017388 & 44.090994 & 0.56 & 0.15 &  $21.60\pm0.03$  & 0.26 & 0.55 &  Young  & \nodata & 70 & \nodata \\
B53  & 187.017608 & 44.090195 & 0.51 & 0.20 &  $21.63\pm0.12$  & 0.30 & 0.59 &  Young  & \nodata & \nodata & \nodata \\
B54  & 187.017134 & 44.089921 & 0.39 & 0.18 &  $21.60\pm0.04$  & 0.21 & 0.63 &  Young  & \nodata & \nodata & \nodata \\
B55  & 187.015220 & 44.087839 & 0.57 & 0.33 &  $21.49\pm0.07$  & 0.16 & 0.42 &  Young  & \nodata & \nodata & \nodata \\
B56  & 187.013513 & 44.088474 & 0.50 & 0.33 &  $22.18\pm0.06$  & 0.14 & 0.31 &  Young  & \nodata & \nodata & \nodata \\
B57  & 187.024797 & 44.093148 & 0.54 & 0.25 &  $22.11\pm0.05$  & 0.57 & 1.10 &  Old  & \nodata & \nodata & \nodata \\
B58  & 187.024237 & 44.092239 & 0.72 & 0.19 &  $22.09\pm0.10$  & 0.14 & 0.70 &  Young  & \nodata & \nodata & \nodata \\
B59  & 187.011491 & 44.084691 & 0.85 & 0.29 &  $20.79\pm0.03$  & 0.56 & 0.96 &  Old  & 4 & \nodata & 129 \\
B60  & 186.981411 & 44.087019 & 1.19 & 0.68 &  $21.85\pm0.09$  & 0.67 & 0.96 &  Old  & \nodata & \nodata & \nodata \\
\hline
\multicolumn{12}{|c|}{Possible Cluster Candidates} \\
\hline
A30  & 187.096924 & 44.109131 & 0.81 & 0.26 &  $22.60\pm0.04$  & 0.57 & 0.88 &  Young  &  \nodata  &  \nodata & \nodata \\
A31  & 187.078726 & 44.091097 & 1.06 & 0.37 &  $21.84\pm0.25$  & 0.76 & 1.21 &  Old       &  \nodata  &  \nodata & \nodata \\
A32  & 187.059613 & 44.091837 & 0.61 & 0.46 &  $20.97\pm0.07$  & 0.70 & 1.08 &  Old       &  \nodata  &  \nodata & \nodata \\
A35  & 187.061156 & 44.094792 & 0.47 & 0.17 &  $20.94\pm0.17$  & 0.12 & 0.45 &  Young  &  \nodata  &  \nodata & 30 \\
A36  & 187.060671 & 44.095046 & 0.54 & 0.26 &  $20.45\pm0.12$  & 0.17 & 0.48 &  Young  &  \nodata  & 25          & \nodata \\
A37  & 187.060062 & 44.094855 & 0.47 & 0.18 &  $20.66\pm0.11$  & 0.15 & 0.62 &  Young  &  \nodata  & 29          & 31 \\
A38  & 187.059857 & 44.093544 & 0.46 & 0.20 &  $20.82\pm0.03$  & 0.26 & 0.55 &  Young  &  \nodata  & 30          & 19 \\
A39  & 187.048740 & 44.091028 & 0.71 & 0.18 &  $20.72\pm0.10$  & 0.30 & 0.73 &  Young  &  \nodata  & 44          & \nodata \\
A40  & 187.052435 & 44.093997 & 0.60 & 0.25 &  $20.17\pm0.22$  & 0.67 & 1.03 &  Old       &   \nodata  &  \nodata & \nodata \\
A41  & 187.053171 & 44.097127 & 0.50 & 0.26 &  $19.61\pm0.13$  & 0.34 & 0.61 &  Young  &  \nodata  & 32           & 44 \\
A42  & 187.052256 & 44.104272 & 0.63 & 0.14 &  $21.11\pm0.48$  & 0.37 & 0.71 &  Young  &  \nodata  &  \nodata & \nodata \\
A44  & 187.036388 & 44.097892 & 0.60 & 0.85 &  $21.88\pm0.76$  & 1.05 & 1.14 &  Old       &  \nodata  &  \nodata & \nodata \\
A47  & 187.018794 & 44.123969 & 0.62 & 0.20 &  $22.74\pm0.04$  & 0.86 & 1.02 &  Old       &  \nodata  &  \nodata & \nodata \\
B22  & 187.055416 & 44.088415 & 0.61 & 0.37 &  $20.56\pm0.03$  & 0.26 & 0.64 &  Young  & 44           &  \nodata & \nodata \\
B23  & 187.049958 & 44.082476 & 0.39 & 0.23 &  $21.79\pm0.05$  & 0.26 & 0.58 &  Young  &  \nodata  &  \nodata & \nodata \\
B24  & 187.050691 & 44.086742 & 0.54 & 0.22 &  $21.05\pm0.05$  & 0.24 & 0.54 &  Young  &  \nodata  &            47 & 5 \\
B40  & 187.013529 & 44.078997 & 0.61 & 0.20 &  $21.84\pm0.03$  & 0.31 & 0.75 &  Young  & 6               &  \nodata & \nodata \\
B42  & 187.004670 & 44.076894 & 0.65 & 0.25 &  $21.90\pm0.03$  & 0.23 & 0.80 &  Young  &  \nodata  &  \nodata & \nodata \\
\enddata

\tablenotetext{a}{Aperture radius for photometry.} 
\tablenotetext{b}{Direct estimate of half-light radius.} 
\tablenotetext{c}{The age of the cluster from its colors, classified as young ($\la1$ Gyr) or old.} 
\tablenotetext{d}{ID from Gelatt \etal~(2001).} 
\tablenotetext{e}{ID from Annibali \etal~(2011a).} 
\tablenotetext{f}{ID from Rangelov \etal~(2011).} 

\end{deluxetable}

\begin{deluxetable}{lcccccccc}
\tablewidth{0pt}
\tablecaption{Dynamical Parameters
        \label{tab:phot}}
\tablehead{ID & $\sigma_p$ & $\sigma_{\infty}$ & $r_h$\tablenotemark{a} & $c$\tablenotemark{b} & log $M_{vir}$ & $M/L_V$ \\
                          &  (\kms) &  (\kms) & (pc) &  & ($M_{\odot}$) & }

\startdata

A8  &  $9.0\pm0.8$ & $9.0\pm0.8$  & 1.74 & 1.54 & $5.51^{+0.08}_{-0.10}$ & $1.45\pm0.30$   \\
A9  &  $5.5\pm0.8$ & $5.3\pm0.8$  & 4.29 & 1.53 & $5.45^{+0.12}_{-0.17}$  & $1.30\pm0.38$   \\
A22 & $9.7\pm0.7$  & $9.6\pm0.7$  & 2.70 & 1.54 & $5.76^{+0.07}_{-0.08}$  & $1.18\pm0.21$  \\
B13\tablenotemark{c} & $15.2\pm0.7$ & $13.8\pm0.7$ & 7.23 & \nodata & $6.51^{+0.06}_{-0.07}$ & $2.83\pm0.53$  \\ 
B15 & $10.3\pm0.7$ & $10.0\pm0.7$ & 3.30 & 2.16 & $5.89^{+0.07}_{-0.08}$ & $1.14\pm0.20$  \\
S121 & $5.3\pm0.8$ & \nodata & \nodata  & \nodata & \nodata & \nodata \\
\hline
A14\tablenotemark{c} & $11.0\pm1.2$ & $10.1\pm1.2$ & 6.48 & \nodata & $6.18^{+0.11}_{-0.14}$ & \nodata \\
A15 & $22.2\pm0.9$ & $20.5\pm0.8$ & 7.24 & 1.44 & $6.85^{+0.05}_{-0.06}$ & \nodata \\
A21 & $6.6\pm0.5$  & $6.4\pm0.5$  & 3.56 & 1.58 & $5.53^{+0.07}_{-0.09}$ & \nodata \\

\enddata

\tablecomments{The first set of objects are old clusters; the second set are younger. An uncertainty of 10\% in $r_h$ and 15\% in $c$ is assumed.}

\tablenotetext{a}{These values assume a distance of 3.82 Mpc to NGC 4449.} 
\tablenotetext{b}{$c$ = log($r_t$/$r_0$) for tidal radius $r_t$ and King radius $r_0$.} 
\tablenotetext{c}{Because of the poor King model fit for these clusters, we use the direct estimate of $r_h$ (assuming an increased uncertainty of 15\%) and assume $c=1.5$ for the purposes of deriving $\sigma_{\infty}$.)} 

\end{deluxetable}

\begin{deluxetable}{lccll}
\tablewidth{0pt}
\tablecaption{Heliocentric Radial Velocities
        \label{tab:rv}}
\tablehead{ID & R.A.~(J2000) & Dec.~(J2000) & R.~V. & Source\tablenotemark{a} \\
                          & (deg) & (deg) & (\kms)     &         }
 
\startdata

A0 & 187.112493 & 44.116711 & $245.8\pm2.7$ & Chelle \\
A6 & 187.078246 & 44.088684 & $103.5\pm1.5$ & Chelle \\
A8 & 187.060675 & 44.083354 & $201.8\pm1.2$ & Chelle \\
A9 & 187.078096 & 44.106327 & $206.4\pm1.1$ & Chelle \\
A14 & 187.054929 & 44.099943 & $192.0\pm1.2$ & Chelle \\
A15 & 187.046198 & 44.093642 & $203.7\pm1.5$ & Chelle \\
A19 & 187.068282 & 44.124788 & $148.6\pm1.5$ & Chelle \\
A21 & 187.058126 & 44.111920 & $207.1\pm0.7$ & Chelle \\
A22 & 187.042842 & 44.105827 & $225.9\pm0.7$ & Chelle \\
A23 & 187.048364 & 44.122332 & $271.9\pm17.9$ & Spec \\
A25 & 187.056230 & 44.143925 & $182.1\pm2.4$ & Chelle \\
A27 & 187.030319 & 44.122556 & $244.1\pm24.2$ & Spec \\
B6 & 187.032557 & 44.084227 & $201.1\pm2.2$ & Chelle \\
B8 & 187.019341 & 44.081410 & $154.9\pm1.2$ & Chelle \\
B9 & 187.015970 & 44.070817 & $107.7\pm1.0$ & Chelle \\
B13 & 186.999279 & 44.078597 & $248.6\pm0.7$ & Chelle \\
B15 & 186.989519 & 44.091054 & $189.0\pm0.6$ & Chelle \\
B16 & 186.970308 & 44.081423 & $170.0\pm2.8$ & Chelle \\
S59 & 186.985667 & 44.096750 & $224.5\pm3.2$ & Chelle \\
S73 & 186.997542 & 44.129667 & $106.5\pm20.8$ & Spec \\
S115 & 186.972292 & 44.075000 & $299.4\pm51.0$ & Spec \\
S117 & 187.114208 & 44.044611 & $226.1\pm2.2$ & Chelle \\
S121 & 187.131083 & 44.046972 & $191.2\pm0.9$ & Chelle \\
S125 & 187.033167 & 44.022278 & $226.3\pm2.4$ & Chelle \\
S162 & 186.987750 & 44.150083 & $106.8\pm29.3$ & Spec \\
S166 & 187.046667 & 44.157444 & $213.1\pm20.0$ & Spec \\
S186 & 186.950667 & 44.048806 & $275.1\pm2.3$ & Chelle \\
S212 & 187.098625 & 43.987500 & $172.0\pm37.2$ & Spec \\
S267 & 187.010167 & 44.179917 & $204.6\pm21.0$ & Spec \\
S464 & 186.946333 & 44.201361 & $139.9\pm1.7$ & Chelle \\
S920 & 187.206167 & 44.217278 & $143.4\pm32.3$ & Spec \\

\enddata

\tablenotetext{a}{Denotes whether the radial velocity is from Hectochelle (Chelle) or Hectospec (Spec).}

\end{deluxetable}

\begin{deluxetable}{lcccccccc}
\tablewidth{0pt}
\tablecaption{SDSS Candidate Globular Clusters with Spectroscopy
        \label{tab:phot_vel}}
\tablehead{ID & R.A.~(J2000) & Dec.~(J2000) & $u$ & $g$ & $r$ & $i$ & $z$ & Class\tablenotemark{a} \\
                          &  (deg)               & (deg)       & (mag) &  (mag) & (mag) & (mag) & (mag)}
 
\startdata

S117 & 187.114216 & 44.044622 & $21.45\pm0.15$ & $19.98\pm0.03$ & $19.44\pm0.03$ & $19.20\pm0.03$ & $19.10\pm0.05$ & 1 \\
S121 & 187.131090 & 44.046976 & $20.85\pm0.09$ & $19.59\pm0.02$ & $18.96\pm0.03$ & $18.71\pm0.02$ & $18.56\pm0.03$ & 1 \\
S267 & 187.010191 & 44.179933 & $21.60\pm0.14$ & $20.21\pm0.03$ & $19.68\pm0.02$ & $19.40\pm0.02$ & $19.24\pm0.05$ & 1 \\
S125 & 187.033167 & 44.022290 & $22.03\pm0.18$ & $20.55\pm0.04$ & $19.94\pm0.03$ & $19.70\pm0.03$ & $19.53\pm0.07$ & 2 \\
S166 & 187.046694 & 44.157450 & $20.91\pm0.08$ & $19.55\pm0.03$ & $19.13\pm0.02$ & $18.90\pm0.02$ & $18.78\pm0.04$ & 3 \\
S186 & 186.950685 & 44.048824 & $21.14\pm0.11$ & $20.02\pm0.03$ & $19.51\pm0.04$ & $19.29\pm0.02$ & $19.23\pm0.06$ & 3 \\
S212 & 187.098625 & 43.987530 & $21.46\pm0.22$ & $20.82\pm0.04$ & $20.30\pm0.04$ & $20.15\pm0.04$ & $19.92\pm0.10$ & 3 \\
S464 & 186.946356 & 44.201382 & $21.76\pm0.16$ & $20.64\pm0.03$ & $20.10\pm0.02$ & $20.00\pm0.03$ & $19.81\pm0.07$ & 3 \\
S73 &  186.997547 & 44.129681  & $20.45\pm0.06$ & $19.41\pm0.02$ & $18.95\pm0.03$ & $18.76\pm0.02$ & $18.74\pm0.04$ & 4 \\
S162 & 186.987768 & 44.150082 & $20.76\pm0.07$ & $19.84\pm0.02$ & $19.68\pm0.02$ & $19.59\pm0.02$ & $19.57\pm0.06$ & 4 \\
S920 & 187.206198 & 44.217285 & $21.03\pm0.09$ & $20.10\pm0.03$ & $19.78\pm0.02$ & $19.80\pm0.04$ & $19.70\pm0.07$ & 4 \\
S59 &  186.985671 & 44.096773  & $20.57\pm0.10$ & $20.66\pm0.07$ & $20.03\pm0.04$ & $20.68\pm0.08$ & $21.02\pm0.26$ & 5 \\
S115 & 186.972306 & 44.075021 & $19.97\pm0.05$ & $20.11\pm0.03$ & $20.28\pm0.04$ & $20.44\pm0.05$ & $20.49\pm0.16$ & 5 \\
\enddata

\tablenotetext{a}{Classification of the candidate. 1: confirmed globular cluster, 2: probable globular cluster, 3: possible globular cluster, 4: foreground star or young cluster, 5: young cluster. These classifications are described in more detail in the text.} 

\end{deluxetable}

\begin{deluxetable}{lccccccc}
\tablewidth{0pt}
\tablecaption{Additional Photometric Candidate SDSS Globular Clusters
        \label{tab:phot_cand}}
\tablehead{ID & R.A.~(J2000) & Dec.~(J2000) & $u$ & $g$ & $r$ & $i$ & $z$  \\
                          &  (deg)               & (deg)       & (mag) &  (mag) & (mag) & (mag) & (mag)}
                          
\startdata
SC1 & 187.106497 & 43.996951 & $21.27\pm0.13$ & $20.27\pm0.03$ & $19.65\pm0.03$ & $19.48\pm0.03$ & $19.28\pm0.06$ \\
\hline
SC2 & 186.894750 & 44.214878 & $20.21\pm0.05$ & $18.84\pm0.02$ & $18.31\pm0.02$ & $18.14\pm0.02$ & $18.11\pm0.02$ \\
SC3 & 187.032983 & 44.140312 & $21.05\pm0.11$ & $19.96\pm0.03$ & $19.35\pm0.03$ & $19.11\pm0.03$ & $18.98\pm0.04$ \\
SC4 & 187.060998 & 44.048861 & $21.87\pm0.21$ & $20.37\pm0.03$ & $19.71\pm0.03$ & $19.45\pm0.03$ & $19.31\pm0.06$ \\
SC5 & 187.130462 & 44.042027 & $21.64\pm0.17$ & $20.63\pm0.04$ & $19.99\pm0.03$ & $19.70\pm0.03$ & $19.49\pm0.07$ \\
SC6 & 187.200516 & 44.100124 & $22.13\pm0.21$ & $20.75\pm0.03$ & $20.15\pm0.03$ & $19.93\pm0.03$ & $19.63\pm0.07$ \\
SC7 & 187.115623 & 44.099028 & $22.62\pm0.40$ & $20.77\pm0.04$ & $20.14\pm0.03$ & $19.84\pm0.03$ & $19.77\pm0.08$ \\
\enddata

\tablecomments{SC1 is confirmed from HST data; the other objects are photometric candidates.}

\end{deluxetable}

\begin{deluxetable}{lccccccccc}
\tablewidth{0pt}
\tablecaption{Spectroscopic Lick Indices
        \label{tab:phot}}
\tablehead{ID & H$\delta_A$ &  H$\delta_F$ & CN$_1$ & CN$_2$ & Ca4227 & G4300 & H$\gamma_A$ &  H$\gamma_F$ & Fe4383 \\
                          &  (\AA) & (\AA) & (mag) & (mag) & (\AA) & (\AA) & (\AA) & (\AA) & (\AA)}

\startdata

B13 & $1.16\pm0.08$ & $1.59\pm0.05$ & $ -0.029\pm0.002$ & $-0.003\pm0.003$ & $0.56\pm0.05$ & $2.69\pm0.08$ & $-0.41\pm0.08$ & $1.12\pm0.05$ & $1.53\pm0.13$ \\
B15 & $0.82\pm0.10$ & $1.61\pm0.07$ & $ -0.021\pm0.003$ & $ 0.005\pm0.003$ & $0.53\pm0.06$ & $2.58\pm0.10$ & $-0.76\pm0.10$ & $0.83\pm0.07$ & $1.91\pm0.16$ \\
\hline
\vspace{0.5mm}
         & Ca4455 & Fe4531 & C4668 & H$\beta$ & Fe5015 & Mg$_1$ & Mg$_2$ & Mg$b$ &  Fe5270 \\
         & (\AA)       & (\AA)       & (\AA)       & (\AA)       & (\AA)       & (mag)       & (mag)       & (\AA)       & (\AA)   \\
\hline
B13 & $0.48\pm0.07$ & $2.34\pm0.10$ & $0.95\pm0.15$ & $2.24\pm0.06$ & $2.44\pm0.14$ & $0.031\pm0.001$ & $0.062\pm0.002$ & $0.99\pm0.07$ & $1.41\pm0.08$ \\
B15 & $0.62\pm0.08	$ & $1.70\pm0.12$ & $0.72\pm0.19$ & $1.90\pm0.08$ & $2.87\pm0.17$ & $0.031\pm0.002$ & $0.059\pm0.002$ & $0.92\pm0.09	$ & $1.63\pm0.09$ \\
\hline
\vspace{0.5mm}
         & Fe5335 &  Fe5406 &   Fe5709 & Fe5782 & NaD & TiO$_1$ & TiO$_2$ & & \\
         & (\AA)       & (\AA)      & (\AA)       & (\AA)       & (\AA)       & (mag)       & (mag)       &  &   \\
\hline
B13 & $1.31\pm0.09$ & $0.72\pm0.07$ & $0.43\pm0.06$ & $0.34\pm0.06$ & $1.45\pm0.07$ & $0.014\pm0.002$ & $0.021\pm0.001$ \\
B15 & $1.29\pm0.11$ & $0.69\pm0.08$ & $0.38\pm0.07$ & $0.21\pm0.07$ & $1.37\pm0.09$	& $0.008\pm0.002$ & $0.015\pm0.002$ \\
\enddata

\end{deluxetable}

\begin{deluxetable}{lccccccc}
\tablewidth{0pt}
\tablecaption{Cluster Photometry from SDSS and Spitzer/IRAC
        \label{tab:phot}}
\tablehead{ID & $u$ & $g$ & $r$ & $i$ & $z$ & 3.6$\mu$m & 4.5$\mu$m \\
                          &  (mag) &  (mag) & (mag) & (mag) & (mag) & (mag) & (mag)}

\startdata

B13 & $19.52\pm0.05$ & $18.09\pm0.01$ & $ 17.47\pm0.01$ & $17.19\pm0.01$ & $17.07\pm0.01$ & $17.87\pm0.03$ & $18.35\pm0.05$ \\
B15 & $20.04\pm0.07$  & $18.50\pm0.01$ & $17.89\pm0.01$ & $17.62\pm0.01$ & $17.51\pm0.02$ & $18.39\pm0.03$ & $18.80\pm0.05$ \\

\enddata

\tablecomments{The Spitzer/IRAC magnitudes are on the AB system.}

\end{deluxetable}

\end{document}